\documentclass[12pt]{article}

\usepackage{epsfig,amsmath,amssymb,graphics,verbatim,amsfonts,subfigure,psfrag,color}
\usepackage[letterpaper,margin=1.2in]{geometry}

\newtheorem{thm}{Theorem}[section]
\newtheorem{cor}[thm]{Corollary}
\newtheorem{prop}[thm]{Proposition}

\def\R{\mathbb{R}}

\def\N{\mathbb{N}}
\def\P{\mathbb{P}}
\def\Z{\mathbb{Z}}
\def\E{\mathbb{E}}
\def\I{\infty}

\def\txtd{{\textnormal{d}}}
\def\txte{{\textnormal{e}}}
\def\txti{{\textnormal{i}}}

\def\txtD{{\textnormal{D}}}

\newcommand{\be}{\begin{equation}}
\newcommand{\ee}{\end{equation}}
\newcommand{\bea}{\begin{eqnarray}}
\newcommand{\eea}{\end{eqnarray}}
\newcommand{\beann}{\begin{eqnarray*}}
\newcommand{\eeann}{\end{eqnarray*}}
\newcommand{\benn}{\begin{equation*}}
\newcommand{\eenn}{\end{equation*}}

\def\ra{\rightarrow}
\def\I{\infty}

\newcommand{\cF}{{\mathcal F}}  
\newcommand{\cI}{{\mathcal I}}  
\newcommand{\cO}{{\mathcal O}}  
\newcommand{\cS}{{\mathcal S}}  

\begin{document}

\author{Karna Gowda\thanks{Engineering Sciences and Applied Mathematics, 
Northwestern University, Evanston, IL 60208-3125, USA}~ 
and Christian Kuehn\thanks{Institute for Analysis and Scientific Computing, 
Vienna University of Technology, 1040 Vienna, Austria}~\thanks{equal contribution}
}
 
\title{Early-Warning Signs for Pattern-Formation in Stochastic Partial Differential Equations}

\maketitle

\begin{abstract}
There have been significant recent advances in our understanding of the potential use and limitations of early-warning
signs for predicting drastic changes, so called critical transitions or tipping points, in dynamical systems. A focus
of mathematical modeling and analysis has been on stochastic ordinary differential equations, where generic statistical
early-warning signs can be identified near bifurcation-induced tipping points. In this paper, we outline some basic
steps to extend this theory to stochastic partial differential equations with a focus on analytically characterizing
basic scaling laws for linear SPDEs and comparing the results to numerical simulations of fully nonlinear problems. In
particular, we study stochastic versions of the Swift-Hohenberg and Ginzburg-Landau equations. We derive a scaling law
of the covariance operator in a regime where linearization is expected to be a good approximation for the local
fluctuations around deterministic steady states. We compare these results to direct numerical simulation, and study the
influence of noise level, noise color, distance to bifurcation and domain size on early-warning signs.
\end{abstract}

\section{Introduction}
\label{sec:intro}

Drastic sudden changes in dynamical systems, so-called critical transitions or tipping points, occur in a wide variety
of applications. It is often desirable to find early-warning signs to anticipate transitions in order to avoid or
mitigate their effects \cite{SchefferCarpenter}. There has been tremendous recent progress in determining potential
warning signs in various sciences such as ecology \cite{SeekellCarpenterPace,vanNesScheffer}, climate science
\cite{Lenton,Lentonetal}, engineering \cite{CotillaSanchezHinesDanforth,LimEpureanu}, epidemiology
\cite{KuehnZschalerGross,OReganDrake}, biomedical applications \cite{MeiselKuehn,Venegasetal} and social networks
\cite{KuehnMartensRomero}; see also \cite{Schefferetal,Schefferetal2} for concise overviews. For a large class of
critical transitions, the underlying dynamical mechanism involves a slow drift of a system parameter towards a local
bifurcation point, where a fast transition occurs \cite{KuehnCT1}. This class has been referred to as ``B-tipping'' in
\cite{AshwinWieczorekVitoloCox}. A detailed mathematical analysis of the underlying stochastic fast-slow systems,
including their generic scaling laws, can be found in \cite{KuehnCT2}; see also \cite{BerglundGentz} for further
mathematical background.

An example of a warning sign occurring in many stochastic systems is an increase in variance as a bifurcation point is
approached~\cite{CarpenterBrock}. This effect is intrinsically generated by critical slowing down (or ``intermittency''
\cite{GH,Strogatz}), {i.e.}~the underlying deterministic dynamics becoming less stable near the bifurcation point.
Hence, (additive) stochastic fluctuations become dominant approaching a B-tipping point.\medskip

A substantial effort has been made to extract early-warning signs, such as slowing down and variance increase, from
univariate time series {e.g.}~using various time series analysis methods \cite{HeldKleinen,Lentonetal1,LivinaLenton},
normal forms \cite{ThompsonSieber2}, topological methods \cite{BerwaldGidea} and generalized models \cite{LadeGross}.
Although theoretical tests and models with sufficiently large data sets tend to work very well
\cite{Dakosetal3,KuehnCT2}, there are clear limits to predictability \cite{BoettingerHastings}, particularly when
relatively sparse data sets are considered
\cite{CimatoribusDrijfhoutLivinavanderSchrier,Dakosetal,DitlevsenJohnsen,Kwasniok}.\medskip

For systems with spatio-temporal dynamics (and associated spatio-temporal data), the additional data in the spatial
direction may be used to improve existing early-warning signs and to discover new ones. If a system is initialized in a
spatially patterned state instead of a homogeneous one, then measures of the pattern could be considered as
potential candidates to provide warning signs. For example, in \cite{Kefietal} the patchiness of states in a vegetation
model is used. However, for a uniform homogeneous steady state that undergoes a bifurcation, such warning signs are not
expected to be available.\medskip

Many early-warning signs computed for univariate time series have multivariate time series analogs, such as spatial
variance and skewness \cite{DonangeloFortDakosSchefferNes,GuttalJayaprakash1} as well as slowing down and spatial
correlation \cite{Dakosetal2}. An ``averaging'' over the spatial direction, {e.g.}~in the sense of the Moran
coefficient \cite{Dakosetal1}, can be helpful to facilitate direct comparisons with univariate indicators. Also, a
natural alternative to avoid the full complexity of spatio-temporal pattern formation is to focus on early-warning
signs for traveling waves \cite{KuehnFKPP}. Despite these exploratory works, it is quite clear at this point that the
full mathematical analysis of early-warning signs for stochastic spatio-temporal systems is largely uncharted
territory. Furthermore, a better theoretical understanding of spatio-temporal warning signs will significantly
improve practical multivariate time series analysis, which is one of the main motivations for this study.\medskip

For finite-dimensional B-tipping, a quite robust classification scheme \cite{AshwinWieczorekVitoloCox,KuehnCT1} has
been formulated and associated warning signs have been investigated (up to generic codimension-two bifurcations) based
upon normal forms, fast-slow systems and stochastic analysis \cite{KuehnCT2}. Such a detailed scheme is much more
difficult to develop for spatio-temporal systems since there is no complete generic bifurcation theory for all
spatio-temporal systems available. However, it is expected that certain subclasses of stochastic partial differential
equations (SPDEs) have warning-signs near pattern-forming bifurcations that can be studied in detail. For deterministic
partial differential equations (PDEs), quite a number of bifurcations leading to pattern-formation are well studied;
for example, see \cite{CrossGreenside,CrossHohenberg,Hoyle} and references therein.\medskip

Translating and extending the \emph{qualitative} pattern-forming results from PDEs to SPDEs is an extremely active area
of current research. We refer to \cite{Bloemker,GarciaOjalvoSancho} for additional background and references. However,
when searching for early-warning signs, it is important to augment the qualitative results with \emph{quantitative} scaling
laws.\medskip

There has been a lot of interest recently in early-warning signs for spatio-temporal systems\footnote{For
example, at the two recent workshops: (I) ``Critical Transitions in Complex Systems'' at Imperial College London, 19
March--23 March, 2012; (II) ``Tipping points: fundamentals and applications'' at ICMS Edinburgh, 9 September--13
September, 2013.}. Furthermore, early-warning signs for particular models have been studied in the context of modeling
case studies. For example, measures of one-point temporal variance and correlation for spatio-temporal processes, which
track temporal statistics at one spatial point, are natural extensions of the generic early-warning signs developed for
univariate time series~\cite{Schefferetal}. Previous studies have computed these measures in moving windows for various
two-dimensional (2D) processes generated by simulations in which a parameter drifts slowly in time, finding that
signals of impending transitions are sometimes obscured by past measurements in the moving
window~\cite{DonangeloFortDakosSchefferNes, Dakosetal1}. For a process generated by a pattern-forming vegetation model,
however, autocorrelation at lag 1 is found to increase monotonically approaching a Turing
bifurcation~\cite{Dakosetal2}.\medskip

More robust signals of critical transitions in spatio-temporal processes are expected to lie in explicitly spatial
measures. Such measures computed at points in time, in contrast with temporal measures computed in a moving window,
reflect the instantaneous (rather than residual) state of a system~\cite{GuttalJayaprakash1}. Spatial variance,
skewness, correlation length, and patchiness have previously been shown to increase before a sudden transition for
various 2D processes~\cite{GuttalJayaprakash1, DonangeloFortDakosSchefferNes, Dakosetal1, Dakosetal2}.\medskip

Although it may seem intuitively clear that the classical warning signs from SODEs should also be found in SPDEs on
bounded domains, there is no complete mathematical theory available to address how classical early-warning signs can
be generalized from stochastic ordinary differential equations (SODEs) to SPDEs. In this paper, we limit ourselves to
several elementary steps working toward this generalization:

\begin{itemize} 
	\item[(R1)] We review the available literature from various fields. In particular, there are closely related
contributions from statistical physics, dynamical systems, stochastic analysis, theoretical ecology and numerical
analysis.
	\item[(R2)] We outline the basic steps to generalize classical SODE warning signs, such as autocorrelation and
variance increase, to the spatio-temporal setting motivated by two standard models for pattern formation, the
Swift-Hohenberg (SH) equation and the Ginzburg-Landau (GL) equation. In particular, we focus on a regime before the
bifurcation, where the linearization around the homogeneous state is expected to provide a very good
approximation to local stochastic fluctuations. The main result is a scaling law of the covariance operator before
bifurcation from a homogeneous branch.
	\item[(R3)] We numerically investigate the SH and GL equation to connect back to several spatio-temporal warning
signs proposed in applications, particularly in the context of ecological models. We compare the
numerical results for the nonlinear systems to the analytical results obtained from linear approximation in (R2). The
numerical results reveal two distinct scaling regimes. Furthermore, we obtain several additional numerical results
about the influence of several natural parameters (domain size, distance to bifurcation, noise level and noise
correlation length) on early-warning signs.

\end{itemize}

Our main results in (R2)-(R3) clearly show that for SPDEs on bounded spatial domains, the classical results from SODEs
are expected to carry over for large classes of SPDEs. In addition, the calculation we carry out in (R2) for linear
SPDEs works directly on the level of covariance operators without using any preliminary dimension reduction techniques.
The numerical simulation results (R3) provide immediate understanding on the influence of many practical parameters of
the problem so that our results are more directly applicable in the analysis of multivariate
time series arising from spatial data. Of course, in the nonlinear regime very close to the bifurcation point,
additional analysis will be necessary and the present study only presents a numerical simulation approach to this
problem.\medskip

This paper is structured as follows: In Section \ref{sec:setup}, we summarize the relevant background for the SH and GL
PDEs. We also recall some basic techniques for studying SPDEs, with a focus on the relationship between correlation
functions and Q-Wiener processes used to define the spatio-temporal noise. In Section \ref{sec:warning_lin}, we
analytically investigate the covariance operator of the linearized SPDE to capture behavior in the regime where we
expect to observe the first warning signs of an approaching bifurcation-point. Here, we find a very natural
generalization to SPDEs of the SODE variance-increase as an early-warning sign. In Section
\ref{sec:numerical_investigation}, we carry out a numerical investigation of SPDE
early-warning signs. In particular, we use numerical simulation results of the SH and GL SPDEs to illustrate the
analytical scaling laws and to link our results to several warning signs proposed recently in applications. Another
purpose for the numerical simulations is to understand the influence of several system parameters. Section
\ref{sec:conclusions} provides a brief outlook of possible future work. \ref{ap:numerics} contains an overview of the
numerical methods we used.

\section{The setup}
\label{sec:setup}

In this section, we review the background required for the subsequent sections. For the review of deterministic PDEs,
we briefly state the main results we need in this paper. The theory needed for our example PDEs, the SH and GL
equations, is quite well-established. The stochastic analysis is less well-known, and we shall thus explain a bit more
for the review of SPDEs.

\subsection{The deterministic PDE(s)} \label{ssec:det_PDE}

We focus on the one-dimensional spatial case on a bounded domain and use the notation 
\benn 
	\cI:=[0,L], \qquad (x,t)\in\cI\times [0,T]\qquad \text{and}\qquad u=u(x,t)\in\R. 
\eenn Let $X$ be a (complex) Hilbert space with inner
product denoted by $\langle \cdot,\cdot\rangle$. Let $A_{(\cdot)}:D(A_{(\cdot)})\subset X\ra X$ be a linear operator
with a dense domain $D(A_{(\cdot)})$. Assume $A_{(\cdot)}$ is the infinitesimal generator of a strongly continuous
semigroup $\txte^{tA}$ \cite{Pazy}; the subscript, indicated via the placeholder $(\cdot)$ distinguishes the two
differential operators we consider below. For functions $v:\cI\ra \R$, we use the standard notation for $L^p$ spaces
with the norms
\benn 
	\|v\|_{L^{p}(\cI)}^p:=\int_\cI|v(y)|^p~\txtd y, 
\eenn 
for $p\in[1,\I)$. Mirroring the finite-dimensional classification scheme for B-tipping \cite{KuehnCT2}, we specify a
class of deterministic systems of the form
\be 
	\label{eq:PDE} \partial_t u = Au+f(u)=ru + A_{(\cdot)}u + f(u),\qquad 
\ee 
where $f(u)=f(u(x,t))$ is a sufficiently smooth polynomial nonlinearity with $f:\R\ra \R$,
$A:=A_{(\cdot)}+r~\textnormal{Id}$ is a linear operator and we use the shorthand notation
$\partial_t=\frac{\partial}{\partial t}$.\medskip

The primary example of \eqref{eq:PDE} we consider in this paper is the Swift-Hohenberg (SH) equation
\cite{SwiftHohenberg} 
\be 
	\label{eq:SH_PDE} \partial_t u = r_{\textnormal{SH}}u - (1+\partial^2_{x})^2u - u^3, 
\ee
where $r_{\textnormal{SH}}\in\R$ is a parameter and $A_{\textnormal{SH}}:=-(1+\partial^2_{x})^2$ is defined on a
suitable domain that is dense in the Hilbert space $X=\textnormal{L}^2(\cI)$ (we employ the notation
$\partial^2_{x}=\frac{\partial^2}{\partial x^2}$). It can be verified that $A_{\textnormal{SH}}$ generates an analytic
semigroup under mild conditions. For periodic boundary conditions, there is a convenient set of orthonormal
eigenfunctions of $A_{\textnormal{SH}}$ given by $a^\textnormal{SH}_k(x)=\frac1L\exp\left(\frac{2\pi}{L}\txti
kx\right)$ for $k\in\Z$ with associated eigenvalues
\benn 
	\lambda^{\textnormal{SH}}_k=-\left(1-\frac{4 \pi^2 k^2}{L^2}\right)^2 \qquad \text{for $k\in\Z$.} 
\eenn 
For simplicity, we consider $L=2\pi$, which yields to the eigenvalues $\lambda^{\textnormal{SH}}_0=-1$, $\lambda^{\textnormal{SH}}_k=-(1-k^2)^2$,
and $k=\pm1$ yield elements in $\text{nullspace}(A_\textnormal{SH})$. For $k\neq \pm1$, we have
$\lambda^{\textnormal{SH}}_k\leq-1$. Linearizing \eqref{eq:SH_PDE} around the trivial solution $u\equiv 0$, leads to
the linear problem
\benn
	\partial_tU=(r_{\textnormal{SH}}\textnormal{Id}+A_{\textnormal{SH}})U=AU,\qquad U\in X. 
\eenn 
Hence, we observe that $u\equiv 0$ is linearly stable for $r_{\textnormal{SH}}<0$ and that a bifurcation occurs at
$r_{\textnormal{SH}}=0$. For a detailed bifurcation analysis of the SH equation we refer to
\cite{BurkeKnobloch,ColletEckmann,CrossHohenberg} and references therein. When we consider a stochastic version of
\eqref{eq:SH_PDE} below, we focus on the parameter regime $r_{\textnormal{SH}}\in[-r_0,0)$ for some $r_0>0$, since it
is our goal to find early-warning signs before the bifurcation occurs.\medskip

As a second example, we study the real Ginzburg-Landau (GL) equation \cite{ColletEckmann1,vanHarten}, which can also be
written in the form \eqref{eq:PDE}. It is given by 
\be 
	\label{eq:GL_PDE} \partial_t u = r_{\textnormal{GL}}u + \partial^2_{x}u - u^3 = r_{\textnormal{GL}}u + A_{\textnormal{GL}}u - u^3, 
\ee 
where $r_{\textnormal{GL}}\in\R$ is a parameter. The Laplacian $\partial_{x}^2$ with periodic boundary conditions on
$[0,L]$ has eigenfunctions
\be
	\label{eq:base_Laplace} \frac{1}{L}\exp\left(\frac{2\pi}{L}\txti k x\right) 
\ee 
with eigenvalues $\lambda_k^{\textnormal{GL}}=-4k^2 \pi^2 /L^2$. Note that, as before, the basis
\eqref{eq:base_Laplace} is orthonormal in $\textnormal{L}^2(\cI)$, and that this general result applies to the
linearization of the GL equation. Hence, the analysis yields that $u\equiv 0$ is linearly stable when
$r_{\textnormal{GL}}<0$ and the first eigenvalue crossing occurs when $r_{\textnormal{GL}}=0$ associated to the
critical eigenvalue $\lambda_0^{\textnormal{GL}}=0$. As in the SH equation, we would like warning signs to predict the
bifurcation point from data obtained in the parameter regime $r_{\textnormal{GL}}\in[-r_0,0)$ for some $r_0>0$.\medskip

We remark that there is a classical connection between the SH and GL equations: the GL equation can be derived as an
amplitude equation of the SH equation \cite{ColletEckmann1,KirrmannSchneiderMielke}. Here, however, we take the view of
studying it independently. The view of GL as an amplitude equation, and the relation between warning signs for the two
models, will be considered in future work.

\subsection{The stochastic PDE(s)} \label{ssec:stoch_PDE}

One efficient approach for obtaining early-warning signs near bifurcation-induced critical transitions is to use
stochastic perturbations to measure the effect of critical slowing down before the bifurcation point, for example
through variance and autocorrelation. For some univariate time series, using ordinary differential equations (ODEs) to
model deterministic dynamics leads quite naturally to SODEs \cite{KuehnCT2}. Following the same paradigm, we search for
warning signs for SPDEs that arise by stochastic perturbations of \eqref{eq:PDE}. Consider SPDEs of the form 
\be
	\label{eq:SPDE} \partial_t u = Au+f(u)+\sigma F(u) \xi(x,t), 
\ee 
where $(x,t)\in\cI\times [0,T]$, $u=u(x,t)$, the maps $f$ and $F$ are assumed to be sufficiently smooth,
$\sigma>0$ controls the noise level and the noise process $\xi(x,t)$ must be specified. Often the noise term is
specified through its correlation function
\be 
	\label{eq:correlation} \E[\xi(x,t)\xi(y,s)]=C_{\textnormal{tem}}(t,s)C_{\textnormal{spa}}(x,y), 
\ee 
where $C_{\textnormal{tem}}$ denotes the temporal correlation function and $C_{\textnormal{spa}}$ the spatial
correlation function; we shall make the assumption \eqref{eq:correlation} throughout this manuscript. As an example,
space-time white noise is given by
\benn
	C_{\textnormal{tem}}(t,s)=\delta(t-s)\qquad \text{and}\qquad C_{\textnormal{spa}}(x,y)=\delta(x-y), 
\eenn 
where $\delta$ denotes the delta-distribution. Although this formulation is quite practical, here the term $\sigma
F(u)\xi(x,t)$ is formal but it can be defined rigorously \cite{Walsh} in certain situations; see also
\cite{Hairer,GubinelliTindel}.\medskip

We now describe one way to provide a rigorous interpretation of \eqref{eq:SPDE} following the approach in
\cite{DaPrato1, DaPratoZabczyk}. We fix a probability space $(\Omega,\cF,\P)$ and let $Q:X\ra X$ be a linear bounded
self-adjoint nonnegative operator on the Hilbert space $X$ with a complete orthonormal set of eigenfunctions
$\{q_k\}_{k=-\I}^\I\subset X$ and associated nonnegative eigenvalues $\{\rho_k\}_{k=-\I}^\I\subseteq [0,\I)$ such that
\be 
	\label{eq:Qbasis} Qq_k=\rho_kq_k\qquad \text{for $k\in\Z$.} 
\ee 
Let $\beta_k(t)$ denote a sequence of independent standard Brownian motions and define the $Q$-Wiener process $W=W(t)$
by
\be 
	\label{eq:noise_expansion} W(t):=\sum_{k=-\I}^\I \sqrt{\rho_k}\beta_k(t)q_k. 
\ee 
If $\textnormal{Tr}(Q)<+\I$ the operator $Q$ is of trace class and the series \eqref{eq:noise_expansion} converges in
$L^2(\Omega,\cF,\P;X)$. If $Q=\textnormal{Id}$ then $\textnormal{Tr}(Q)=+\I$ and $W$ is a cylindrical Wiener process.
If $X_1$ is a Hilbert space into which $X$ continuously embeds and for which the embedding from $X_0:=Q^{1/2}X$ to $X_1$
is Hilbert-Schmidt, then the series \eqref{eq:noise_expansion} converges in $L^2(\Omega,\cF,\P;X_1)$; see also
\cite{DaPratoZabczyk} for more details and the technical complications of cylindrical Wiener processes. We remark that
it is common to index $q_k$ and $\beta_k$ using the natural numbers only, but here it is more convenient to use integer
indices. It is often convenient also to take $X=L^2(\cI)$, as already considered above for the deterministic case. 
We focus on additive noise for \eqref{eq:SPDE} ({i.e.}~when $F(u)$ is constant) and write it in the form
\be 
	\label{eq:DaPrato} \txtd u=(Au+f(u))~\txtd t+\sigma B~\txtd W_t,\qquad u(x,0)=u_0(x), 
\ee
where $W(t)$ is a $Q$-Wiener process, $f$ is a sufficiently smooth map, $B:X\ra X$ is a bounded linear operator, $\sigma>0$ and
$u_0\in X$ is assumed to be deterministic. For our purposes, it will suffice to view \eqref{eq:DaPrato} as an evolution
equation for $u(\cdot,t)=:u(t)$ and formally consider mild solutions \cite[{Ch.7}]{DaPratoZabczyk} given by 
\be
	\label{eq:mild} u(t)=\txte^{tA}u_0+\int_0^t \txte^{(t-s)A}f(u(t))~\txtd s+\sigma \int_0^t\txte^{(t-s)A}B~\txtd W(s),
\ee 
where the stochastic integral with respect to $W(s)$ can be defined as a limit of finite-dimensional approximations
\cite[Sec.~4.3.2.]{DaPratoZabczyk} by truncating the series \eqref{eq:noise_expansion} and using the usual definition
of the stochastic integral with respect to $\beta_k$ \cite{Oksendal}.\medskip

For convenience, we denote the stochastic integral in \eqref{eq:mild} as 
\benn 
	W_A(t)=\int_0^t\txte^{(t-s)A}B~\txtd W(s) 
\eenn 
and refer to it as the stochastic convolution. One of its most important properties is the expression for the
associated covariance operator \cite[{Thm. 5.2}]{DaPratoZabczyk}
\be 
	\label{eq:CovConv} V(t):=\textnormal{Cov}(W_A(t))=\int_0^t \txte^{rA}BQB^*\txte^{rA^*}~\txtd r 
\ee 
where $B^*$ denotes the adjoint operator of $B$ and similarly $\txte^{rA^*}$ denotes the adjoint semigroup of $\txte^{rA}$.\medskip

We remark that once the operators $Q$ and $B$ are fixed, this determines the correlation structure of 
the additive noise as defined in \eqref{eq:correlation}. Indeed, we have for any $g,h\in X$ and $t,s\geq0$ that
\benn
\E[\langle W_{\textnormal{0}}(t),g\rangle \langle W_{\textnormal{0}}(s),h\rangle]
=\min(t,s) \langle BQB^*g,h\rangle,
\eenn
which is equivalent to the more detailed formulation
\benn
\int_\cI \int_\cI\E[ W_{\textnormal{0}}(t)  W_{\textnormal{0}}(s)]
g(x)h(y)~ \txtd x~\txtd y=\min(t,s) \int_\cI Q^{1/2}B^*g(v)Q^{1/2}B^*h(v)~ \txtd v,
\eenn
since $Q$ is self-adjoint. If we consider the basis $q_k$, then we also find
\beann
Q^{1/2}B^*g&=&\sum_{k=-\I}^\I \langle g,q_k\rangle Q^{1/2}B^*q_k
=\sum_{k=-\I}^\I \langle g,q_k\rangle \sum_{l=-\I}^\I \sqrt{\rho_l}\langle B^*q_k,q_l\rangle q_l\\
&=& \sum_{l,k=-\I}^\I \sqrt{\rho_l} \int_\cI g(x)q_k(x)~ \txtd x 
\underbrace{\int_\cI q_k(y)(Bq_l)(y)~\txtd y}_{=:b_{kl}}~ q_l.
\eeann
and similarly for $Q^{1/2}B^*h$. Therefore, it follows that
\beann
&&\int_\cI Q^{1/2}B^*g(v)Q^{1/2}B^*h(v)~ \txtd v\\
&=&\int_\cI \sum_{k,l=-\I}^\I b_{kl}\sqrt{\rho_l}\int_\cI g(x)q_k(x)~\txtd x~ q_l(v)
\sum_{m,n=-\I}^\I b_{mn}\sqrt{\rho_n}\int_\cI h(y)q_m(y)~\txtd y~ q_n(v)~\txtd v \\
&=&\int_\cI \int_\cI \sum_{l=-\I}^\I\rho_l\left(\sum_{k=-\I}^\I  b_{kl}q_k(x)\right)
\left(\sum_{m=-\I}^\I b_{ml}q_m(y)\right)g(x)h(y)~\txtd x~\txtd y\\
&=&\int_\cI \int_\cI \sum_{l=-\I}^\I\rho_l\sum_{n=-\I}^\I c_{nl}(x,y)~ g(x)h(y)~\txtd x~\txtd y,\\
\eeann 
where $c_{nl}$ is computed from the discrete convolution in the usual way
\benn
c_{nl}(x,y)=\sum_{j=-\I}^\I b_{jl}~b_{(n-j)l}~q_j(x) ~q_{n-j}(y).
\eenn
This gives $C_{\textnormal{spa}}(x,y)=\sum_{l=-\I}^\I\rho_l\sum_{n=-\I}^\I c_{nl}(x,y)$, which is colored noise in general. The temporal correlation is white, since differentiating $\min(t,s)$ formally yields a temporal correlation function $C_{\textnormal{tem}}(t-s)=\delta(t-s)$. The relation between the correlation function and a suitable convolution involving $B$ and $Q$ is well-known when $\cI=\R^d$ \cite{BrzezniakPeszat,PeszatZabczyk}.\medskip

To conclude our discussion of SPDEs, we briefly review several works considering stochastic perturbations of PDEs with a focus 
on the SH SPDE. Additive noise, {i.e.}~when $F(u)$ is constant, is considered in \cite{GoldmanSwiftSwinney,
HohenbergSwift} with a comparison to experimental data. Multiplicative noise, {i.e.}~when $F(u)$ depends upon $u$
non-trivially, is studied in \cite{GarciaOjalvoHernandezMachadoSancho} with a focus on noise-induced shifts of the
bifurcation point. Such bifurcation-shifts are also considered in \cite{HuttLongtinSchimansky-Geier1,
HuttLongtinSchimansky-Geier} for additive noise, and in \cite{BeckerKramer} for stochastic variation of the bifurcation
parameter. Parameter fluctuations may also induce stochastic resonance effects in the SH equation \cite{VilarRubi} (for
coherence resonance induced by additive noise, see \cite{CarrilloSantosGarciaOjalvoSancho}). Pattern formation, pattern
selection and convergence to various states in the presence of stochasticity is considered in \cite{ ElderVinalsGrant,
HernandezGarciaSanMiguelToral, VinalsHernandezGarciaSanMiguealToral}. The amplitude equations for the stochastic SH
equation and related models have also been studied extensively in recent years \cite{AgezClercLouvergneauxRojas,
Bloemker1, Bloemker, BloemkerHairerPavliotis1, BloemkerMohammed1, MohammedBloemkerKlepel}. However, there seems to be
relatively little, if any, work yet that focuses on early-warning signs for the stochastic SH equation.

\section{Warning signs from linearization}
\label{sec:warning_lin}

When considering early-warning signs in SODEs perturbed by additive noise, it is very helpful to start with the
analysis around a parametrized curve of attracting steady states of the deterministic system and consider the
approximation by a linear stochastic process. This Ornstein-Uhlenbeck (OU) process has certain growing elements in
its covariance matrix as a bifurcation point is approached \cite{KuehnCT2}. Of course, the regime very close to the
bifurcation is more difficult to study analytically as the nonlinear terms will contribute essential features.
Furthermore, the analysis is complicated by a slow parameter drift in time, see {e.g.}~\cite{BerglundGentz6,
BerglundGentzKuehn} for the SODE case. In this paper, we just treat the first simple step for SPDEs in a regime where
the linear approximation is expected to be a very good local approximation of the dynamics and the parameter drift is
infinitely slow. The full nonlinear regime is considered numerically in Section \ref{sec:numerical_investigation}.

\subsection{Correlation function}
\label{ssec:cor_func}

Consider an SPDE of the form \eqref{eq:SPDE} with $F(u)=\textnormal{Id}$ and $f(u)=0$, {i.e.} a linear SPDE perturbed
by additive noise. We recall a few formal results about the correlation structure of the solution $u(x,t)$ when
$\xi(x,t)$ is space-time white noise, {i.e.}, $Q=\textnormal{Id}$. For $A=r\textnormal{Id}+\partial^2_{x}$, $r<0$,
$\cI=[0,L]$ and periodic boundary conditions, the solution $u(x,t)$ can be written in terms of a Fourier basis and an
associated Green's function \cite{Lythe2} as
\be
\label{eq:sol_SPDE_dom}
u(x,t)=\int_\cI G(x,v,t,0)u_0(v)~\txtd v+\sigma\int_0^t \int_\cI G(x,v,t,s)~\txtd v~ \txtd W(s),
\ee
where $W(s)$ is a cylindrical Wiener process with covariance $Q=\textnormal{Id}$ and the 
Green's function $G$ is given by
\benn
G(x,v,t,s)=\frac{\exp(-2r(t-s))}{2\sqrt{\pi(t-s)}}\sum_{k=-\I}^\I
\exp\left(-\frac{(x-v-kL)^2}{4(t-s)}\right).
\eenn
We define the correlation function of the solution $u(x,t)$ as
\be
\label{eq:corr_func}
c(x,y,s,t):=\E[u(x,t)u(y,s)].
\ee
In formula \eqref{eq:sol_SPDE_dom}, we observe that the first term decays rapidly for 
any initial condition $u_0\in X$ so that the correlation function \eqref{eq:corr_func}
arises primarily from the stochastic integral. By spatial translation invariance, the
correlation function only depends upon the difference $|x-y|$. A leading-order asymptotic 
result obtained in \cite{Lythe2,LytheHabib} is that 
\be
\label{eq:corr_func_simple}
\lim_{t\ra \I}c(x,0,t,t)\sim\frac{\sigma^2}{4\sqrt{|r|}}\txte^{-|x|\sqrt{|r|}}.
\ee
Formulas for the correlation function in higher-dimensions (arising from rather involved calculations) exist
\cite{LamBagayoko,Lythe2,LytheHabib}. The Fourier transform in combination with Bessel potential solutions \cite[{Sec.
4.3}]{Evans} may also be used to calculate the correlation function for the $\cI=\R$ as shown in \cite[{Sec.
2.3}]{Hairer}. However, the formula \eqref{eq:corr_func_simple} suffices here to illustrate that the correlation
function of an SPDE depends in a non-trivial way on the bifurcation parameter. This certainly provides a first hint 
that an SPDE may exhibit early-warning signs before bifurcations. At this point, the formal and asymptotic 
calculations of the correlation function for linear PDEs involving the
Laplacian and a space-time white-noise driving term are already quite complicated. If a more complex space-time
correlation structure is specified via $C_{\textnormal{tem}}$ and $C_{\textnormal{spa}}$, or if a different linear
operator $A$ is chosen, there may be no closed form expression for \eqref{eq:corr_func}. Hence, it appears very useful
to consider an abstract framework to study generic covariance-related early-warning signs.

\subsection{The covariance operator}
\label{ssec:cov_operator}

An alternative approach that does not immediately yield explicit formulas is to use the covariance operator from
\eqref{eq:CovConv}. Suppose $F(u)=\textnormal{Id}$ and $f(u)=0$ so that \eqref{eq:DaPrato} is a linear SPDE with
additive noise and solution given by
\be
\label{eq:formal_sol_DP}
u(x,t)=\txte^{tA}u_0+\sigma W_A(t).
\ee
We will assume that $r<0$ so that 
\benn
\|\txte^{tA}\|\leq M \txte^{\omega t}
\eenn
for some $M>0$ and $\omega<0$. In particular, $\|\txte^{tA}u_0\|_X\ra 0$ as $t\ra +\I$, so we may neglect the first
term $\txte^{tA}u_0$ if we are only interested in the asymptotic behavior as $t\ra +\I$; alternatively, we could set
$u_0(x)= 0$ for all $x\in\cI$. Using \eqref{eq:noise_expansion}, we can now write the solution \eqref{eq:formal_sol_DP}
as the stochastic convolution
\be
\label{eq:formal_series_sol}
u(x,t)=\sigma \sum_{k=-\I}^\I\sqrt{\rho_k}\int_0^t \txte^{(t-s)A}Bq_k~\txtd \beta_k(s)=\sigma W_A(t).
\ee
The result \cite[{Prop. 2.2}]{DaPrato1} requires, aside from the usual strong continuity assumption on $\txte^{tA}$ and
linearity for $B:X\ra X$, that the operator $BB^*$ is of trace class $\textnormal{Tr}(BB^*)<+\I$. Under these
assumptions, the series in \eqref{eq:formal_series_sol} is convergent in $\textnormal{L}^2(\Omega,\cF,\P;X)$. Then it
follows that
\be
\E[\|W_A(t)\|^2]=\textnormal{Tr}(V(t)),\qquad t\geq 0.
\ee
There exists a unique invariant Gaussian measure with mean zero and covariance operator $V_\I:=\lim_{t\ra +\I} V(t)$
\cite[{Thm. 2.34}]{DaPrato1}. $V_\I:H\ra H$ is a linear continuous symmetric operator that satisfies $\langle V_\I
g,g\rangle\geq 0$ for all $g\in X$ and $V_\I=V_\I^*$. Furthermore, $V_\I$ satisfies a Lyapunov equation \cite[{Lem.
2.45}]{DaPrato1} given by
\be
\label{eq:Lyapunov}
\langle AV_\I g,h\rangle +\langle V_\I A^*g, h\rangle=-\sigma^2\langle BQB^* g,h\rangle,
\ee
for all $g,h \in X$, which is a generalization of the classical Lyapunov equation associated with linear SODEs used to determine scaling
laws for warning signs \cite{KuehnCT2}. In fact, a suitable analog of \eqref{eq:Lyapunov} even holds for transition
semigroups in more general Banach spaces \cite[{Sec. 4}]{GoldysvanNeerven}. For SODEs, solving \eqref{eq:Lyapunov}
requires solving a matrix-valued equation, which can not only be solved analytically for certain cases but can also be
efficiently solved numerically for general nonlinear parametrized stochastic systems \cite{KuehnSDEcont1}.\medskip

Solving \eqref{eq:Lyapunov} is more problematic for infinite-dimensional operators. A natural first attempt is to
compute the operator using a suitable basis of $X$. However, there are two natural bases to consider. For one, we can
use the eigenbasis $\{q_k\}_{k=-\I}^\I$ of $Q$ given in \eqref{eq:Qbasis}. Alternatively, we can use
\benn
Aa_k=ra_k+A_{(\cdot)}a_k=(r+\lambda^{(\cdot)}_k)a_k,\qquad a_k\in X, \quad k\in\Z
\eenn
so that $a_k$ are eigenfunctions for $A$ (respectively $A_{(\cdot)}$). In either case, there are now several 
straightforward and instructive calculations we can carry out to understand potential early-warning signs related to $V_\I$.
First, we consider the case in which $B=\textnormal{Id}$ and $a_k=q_k$ for all $k\in \Z$ is an orthonormal basis of
$X$; the operator $A$ has eigenvalues $\lambda_k=r+\lambda_k^{(\cdot)}$ and the operator $Q$ has eigenvalues $\rho_k$
in this basis. The operator $V_\I$ is completely determined if we can compute the coefficients $\langle V_\I
a_k,a_j\rangle$ for all $k,j\in\Z$. From \eqref{eq:Lyapunov}, we find
\beann
0&=&\langle V_\I a_k,A^*a_j\rangle +\langle A^*a_k,V_\I a_j\rangle+\sigma^2\langle Q a_k,a_j\rangle\\
&=& \langle V_\I a_k,\overline{\lambda_j}a_j\rangle +
\langle \overline{\lambda_k}a_k,V_\I a_j\rangle+\sigma^2\langle\rho_k a_k,a_j\rangle\\
&=& \lambda_j\langle V_\I a_k,a_j\rangle +
\overline{\lambda_k}\langle V_\I a_k, a_j\rangle+\sigma^2\rho_k\langle  a_k,a_j\rangle.
\eeann
Using orthonormality of the basis it follows that 
\benn
\langle V_\I a_k,a_j\rangle=\left\{
\begin{array}{ll}
-\sigma^2\frac{\rho_k}{\lambda_k+\overline{\lambda_k}},\quad &\text{if $k=j$,}\\
0,\quad &\text{if $k\neq j$ (and $\lambda_j\neq -\overline{\lambda_k}$).}\\
\end{array}
\right.
\eenn
We note that $\lambda_j \neq -\overline{\lambda_k}$ as long as we have $\textnormal{Re}(\lambda_k)<0$ for all $k\in\Z$.
Therefore, the operator $V_\I$ is diagonal and the important coefficients (for $r<0$) are
\benn
\langle V_\I a_k,a_k\rangle=-\sigma^2\frac{\rho_k}{\lambda_k+\overline{\lambda_k}}=
-\sigma^2\frac{\rho_k}{2r+\lambda_k^{(\cdot)}+\overline{\lambda^{(\cdot)}_k}}=
-\sigma^2\frac{\rho_k}{2\left(r+\textnormal{Re}\left(\lambda_k^{(\cdot)}\right)\right)}\geq 0,
\eenn
where the last inequality follows from $\rho_k\geq 0$ and $r+\textnormal{Re}(\lambda_k^{(\cdot)})< 0$
when $r< 0$. Now we can consider particular eigenvalues $\lambda^{(\cdot)}_k$ for the SH and
GL linearized operators. For example, in the SH equation we have $\lambda^{\textnormal{SH}}_{\pm 1}=0$
as critical eigenvalues. Hence, we find the divergent coefficients
\benn
\lim_{r\ra 0^-}\langle V_\I a_{\pm1},a_{\pm1}\rangle=\lim_{r\ra 0^-}
-\sigma^2\frac{\rho_{\pm 1}}{2r}=+\I,
\eenn
for fixed $\rho_{\pm 1}> 0$. This represents an $\cO(1/r)$ variance scaling law as $r\ra 0^-$
($\sigma>0$ is fixed) for the linearized system, which resembles the variance scaling laws
associated with finite-dimensional bifurcation points ({e.g.}~\cite[{Thm 5.1}]{KuehnCT2}
or \cite{BerglundGentz6}). The same scaling law applies to the GL equation with critical
eigenvalue $\lambda_0^{\textnormal{GL}}=0$.  We remark that if $\rho_{\pm 1}=0$, then no
such scaling law can be expected. Of course, spatio-temporal noise with $\rho_{\pm 1}=0$
is a highly degenerate scenario, and is not expected to occur often in practice.\medskip

Next, we consider the more general case with a linear operator $B$ and in which the orthonormal
eigenbases of $A$ and $Q$ do not coincide. The Lyapunov equation \eqref{eq:Lyapunov} gives that 
\benn
0= \lambda_j\langle V_\I a_k,a_j\rangle +
\overline{\lambda_k}\langle V_\I a_k, a_j\rangle
+\sigma^2\langle BQB^* a_k,a_j\rangle.
\eenn
This implies
\benn
\langle V_\I a_k,a_j\rangle 
=-\sigma^2\frac{\langle BQB^* a_k,a_j\rangle}{\lambda_j+\overline{\lambda_k}}
=-\sigma^2\frac{\langle BQB^* a_k,a_j\rangle}{2r+\lambda^{(\cdot)}_j+\overline{\lambda_k^{(\cdot)}}}.
\eenn
In particular, we have shown the following result:

\begin{prop}
Consider the linear SPDE
\be
\label{eq:DaPrato1}
\txtd u=Au~\txtd t+\sigma B~\txtd W_t,\qquad (x,t)\in\cI\times [0,+\I),~u=u(x,t)
\ee
where $A=r~\text{Id}+A_{(\cdot)}$ and $A_{(\cdot)}$ has a discrete spectrum with eigenvalues $\lambda_k^{(\cdot)}$ with
$\text{Re}(\lambda_k^{(\cdot)})<0$ and eigenfunctions $a_k$. Then the covariance operator $V(t)$ satisfies
\benn
\left\langle \lim_{t\ra +\I}V(t) a_k,a_j\right\rangle =-\sigma^2\frac{\langle BQB^* a_k,a_j\rangle}{2r+\lambda^{(\cdot)}_j+\overline{\lambda_k^{(\cdot)}}}.
\eenn
\end{prop}

If the eigenvalues $\lambda^{(\cdot)}$ are real, as they are for the SH and GL equations 
considered here, it follows that 
\be
\label{eq:covmat_SPDE}
\langle V_\I a_k,a_j\rangle 
=-\sigma^2\frac{\langle BQB^* a_k,a_j\rangle}{2r+\lambda^{(\cdot)}_j+\lambda_k^{(\cdot)}}.
\ee
Note that $V_\I$ is generically non-diagonal, {i.e.},
\benn
\langle BQB^* a_k,a_j\rangle\neq 0 \qquad \text{for all $k,j\in\Z$.}
\eenn
We have already computed the eigenvalues for the SH and GL equations. For example, for the SH equation with $L=2\pi$,
we have
\benn
2r+\lambda^{\textnormal{SH}}_j+\lambda_k^{\textnormal{SH}}=2r-(1-j^2)^2-(1-k^2)^2.
\eenn
Hence if $\langle BQB^* a_k,a_j\rangle\neq 0$ for $k,j=\pm 1$ then a leading-order scaling law for $\langle V_\I
a_k,a_j\rangle$ of order $\cO(1/r)$ as $r\ra 0^-$ is observed for the four coefficient pairs
\benn
(k,j)\in\{(1,1),(1,-1),(-1,1),(-1,1)\}.
\eenn
As another example, we consider the GL equation with $L=2\pi$. The eigenvalues are then $\lambda_k^{\textnormal{GL}}=-
k^2$. Therefore, we find
\be
\label{eq:GL_cov_scale}
\langle V_\I a_k,a_j\rangle 
=-\sigma^2\frac{\langle BQB^* a_k,a_j\rangle}{2r-(k^2+j^2)}.
\ee
and we again observe the $\cO(1/r)$-scaling law for the critical mode when $k=j=0$. If we increase the domain size and
consider $L=2\pi l$ for some $l\in\N$ and $l\gg1$, then \eqref{eq:GL_cov_scale} becomes
\be
\langle V_\I a_k,a_j\rangle 
=-\sigma^2\frac{\langle BQB^* a_k,a_j\rangle}{2r-(k^2+j^2)/l^2}.\label{eq:GL_bigL}
\ee
since the eigenvalues of the Laplacian become $\lambda_k^{\textnormal{GL}}=-k^2/l^2$. Thus, the $\cO(1/r)$ scaling law
begins to appear in all modes if we take the formal limit $l\ra \I$ before considering $r\ra 0^-$. We can summarize the
observations made for the linearized SH and the linearized GL equations in more generality:

\begin{cor}\label{cor:scaling_law}
Consider the linear SPDE
\be
\label{eq:DaPrato2}
\txtd u=Au~\txtd t+\sigma B~\txtd W_t,\qquad (x,t)\in\cI\times [0,+\I),~u=u(x,t)
\ee
where $A=r~\text{Id}+A_{(\cdot)}$ and $A_{(\cdot)}$ has a discrete real spectrum with eigenvalues $\lambda_k^{(\cdot)}$
with $\lambda_k^{(\cdot)}<0$ for $r<0$ and there exists $k^*$ such that $\lambda_{k^*}^{(\cdot)}=0$ for $r=0$. Then the
covariance operator $V(t)$ satisfies
\be
\label{eq:cor_scaling_law}
\left\langle \lim_{t\ra +\I}V(t) a_{k^*},a_{k^*}\right\rangle
=\cO\left(\frac1r\right) \qquad \text{as $r\ra 0^-$}
\ee
if the genericity condition $\langle BQB^* a_{k^*},a_{k^*}\rangle\neq 0$ is satisfied.
\end{cor}

Hence, the scaling law results can be worked out not only for the SH and GL linearized operators but for any SPDE of
the form \eqref{eq:DaPrato} as long as the linear approximation is valid and we bifurcate from a homogeneous state. As
with SODEs, this should be done by linearizing about a steady state of the deterministic system, operating in a regime
below the first destabilizing bifurcation point and using the Lyapunov equation to compute the scaling law for the
associated covariance operator $V_\I$. In particular, the results obtained in \cite{KuehnCT2} are expected to fully
carry over for SPDEs on bounded domains. For example, the scaling law for fold bifurcations at $r=0$ will be
$V_\I=\cO(1/\sqrt{-r})$ as $r\ra 0^-$; we refer also to the recent numerical results in \cite{KuehnSPDEcont}. We also remark that the
scaling law can change if there is a parameter dependence of the eigenvalues $\lambda_k$ and eigenfunctions $a_k$. In
particular, if $\langle BQB^* a_{k^*},a_{k^*}\rangle=\cO(r^\beta)$ and $\lambda_k=\cO(r^\alpha)$ then
\eqref{eq:cor_scaling_law} becomes
\be
	\label{eq:cor_scaling_law1}
	\left\langle V_\I a_{k^*},a_{k^*}\right\rangle = \cO\left(\frac{1}{r^{1-\beta}+2r^{\alpha-\beta}}\right) \qquad \text{as $r\ra 0^-$.}
\ee
Furthermore, we emphasize again that the analytical approach here does not cover the truly nonlinear regime very close
to the bifurcation point, and that a specialized analysis will be necessary for different classes of the nonlinearity;
we discuss scaling laws in the nonlinear regime further in Section \ref{ssec:numerics_cor_scaling_law}.

\section{Numerical investigation of spatio-temporal early-warning signs}
\label{sec:numerical_investigation}

Rather than pursuing the abstract theory further, we proceed in this section by numerically investigating the 
scaling law result from Corollary \ref{cor:scaling_law} in the SH and GL SPDEs. We focus on verifying the existence 
of the scaling law, as well as on identifying the influence of nonlinearity on this law. In addition, we describe 
several generic early-warning signs that have been proposed for spatio-temporal processes. We then explore them 
numerically in the SH SPDE, and take steps toward understanding the influence of several system parameters on the 
computed measures.\medskip

Several previous studies of early-warning signs in spatial systems consider a parameter that drifts slowly in
time~\cite{GuttalJayaprakash1, DonangeloFortDakosSchefferNes, Dakosetal1, Dakosetal2}. Here, we essentially consider
the limit as the parameter drift rate vanishes by simulating the SPDE \eqref{eq:SPDE} using a series of fixed parameter
values. We do this in order to relate the numerical results directly to the analytical results described in Section
\ref{sec:warning_lin}. Furthermore, this approach guarantees well-defined stationary early-warning measures. The 
analysis of a slowly-drifting parameter is postponed for future work.

\subsection{Variance scaling law}
\label{ssec:numerics_cor_scaling_law}

To both verify the scaling law derived in Corollary \ref{cor:scaling_law} and to investigate its regime of validity, we
numerically simulate the GL and SH SPDEs in the form of \eqref{eq:SPDE}, with $F(u) = 1$ and $\sigma = 0.01$. We
take $\xi(x,t)$ as space-time white noise. Simulations are run for values of $r \in [-1,-0.01]$ on domains of
size $L=2\pi$ and $L=16\pi$ with periodic boundary conditions. A spatial finite-difference method was used to
discretize the SPDEs. The resulting SODEs were solved by an implicit Euler-Maruyama method. For a more detailed
description of the numerical method we refer to \ref{sec:fin_diff_spde}.\medskip 

\begin{figure}[htbp]
	\psfrag{a}{(a)} 
	\psfrag{b}{(b)}
	\psfrag{c}{(c)} 
	\psfrag{d}{(d)}
	
	\psfrag{r}[c]{\scriptsize $\log_{10}(-r)$}
	\psfrag{u}[c]{\scriptsize $\log_{10}(V_k)$}
	
	\psfrag{l}[l]{\scriptsize $k=0$}
	\psfrag{m}[l]{\scriptsize $k=1$}
	\psfrag{n}[l]{\scriptsize $k=2$}
	\psfrag{o}[l]{\scriptsize $k=3$}
	\psfrag{p}[l]{\scriptsize $k=8$}
	\psfrag{q}[l]{\scriptsize $k=7$}
	\psfrag{s}[l]{\scriptsize $k=9$}
	\psfrag{t}[l]{\scriptsize $k=10$}
	
	\psfrag{G}[c]{\scriptsize GL, $L=2\pi$}
	\psfrag{S}[c]{\scriptsize SH, $L=2\pi$}
	\psfrag{F}[c]{\scriptsize GL, $L=16\pi$}
	\psfrag{R}[c]{\scriptsize SH, $L=16\pi$}
	
	\centering
	\epsfig{file=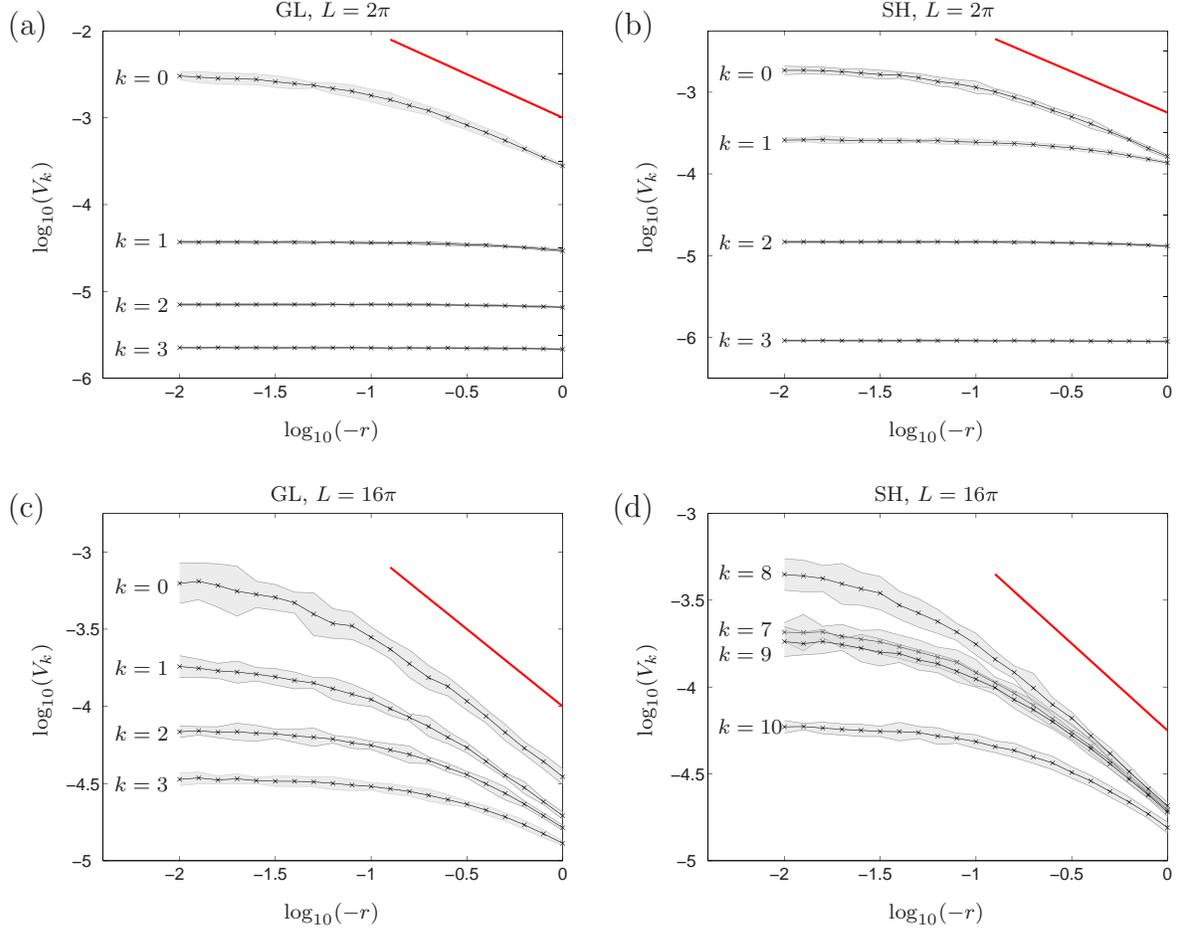,width=1\textwidth}
	\caption{Log-log plot of variances of critical and near-critical Fourier modes \eqref{eq:fourier_variance} in
solutions of the GL (a, c) and SH (b, d) SPDEs, depending on the parameter $r \in [-1,-0.01]$; solutions are computed
using domain sizes $L=2\pi$ (a, b) and $L=16\pi$ (c, d); guide lines proportional to $\log_{10}(-1/r)$ are plotted in
solid red; the mean values of $V_k$ over 10 simulations are plotted in black, and the grey regions show neighborhoods
of three sample standard deviations about these means. Here we have chosen $\xi(x,t)$ to be space-time white noise,
$F(u)=1$ and $\sigma = 0.01$. SPDEs were numerically simulated as described in \ref{sec:fin_diff_spde} .}
\label{fig:fourier_scaling}

\end{figure}

In order to compare numerical solutions $u(x,t;r)$ with the theoretical prediction of a scaling law in the 
covariance operator of the solution, we compute the following measure in Fourier space:
\be
	V_k(r) = \frac{1}{M} \sum_{n=1}^{M} (|\hat{u}(k,t;r)| - \bar{u}(k;r))^2,\label{eq:fourier_variance}
\ee
where $\hat{u}(k,t;r) = \frac{1}{L} \int_{\cI} u(x,t;r) \exp(-2\pi \txti k x/L) ~\txtd x$ and $\bar{u}(k;r) = \frac{1}{M}
\sum_{n=1}^{M} |\hat{u}(k,t_n;r)|$. Here, the eigenfunctions $a_k$ are taken to be Fourier modes and an
$\mathcal{O}(1/r)$ scaling law is expected for the variance in the modes for which $\lambda_k = 0$, {i.e.}~the
critical modes of the GL and SH operators. For simulations in which the domain size is taken to be $L=2\pi$, such
scaling is indeed observed in the critical modes of the SPDEs (at $k=0$ for GL and $k = 1$ for SH). Figure
\ref{fig:fourier_scaling} plots $\log_{10}(-r)$ against $\log_{10}(V_k)$, with guide lines proportional to 
$\log_{10}(-1/r)$. When $r$ is sufficiently far from $r=0$ ({i.e.}~when the linearization is a good approximation), 
these critical Fourier modes appear to follow the predicted scaling. For comparison, the $V_k$ measures for adjacent 
modes are also plotted in Figure \ref{fig:fourier_scaling} and are observed to scale much more slowly than $\cO(1/r)$.\medskip

Another set of predictions from Section \ref{ssec:cov_operator} describes the scaling of variance in non-critical modes
when $L \gg 2\pi$. Equation \eqref{eq:GL_bigL} shows that an $\cO(1/r)$ scaling should begin to appear in near-zero
modes for the GL operator as $L \to \infty$ (where $L = 2\pi l$). Similarly for the SH operator,
\benn
	\langle V_\I a_k,a_j\rangle =-\sigma^2\frac{\langle BQB^* a_k,a_j\rangle}{2r-(1-j^2/l^2)^2 -(1-k^2/l^2)^2},
\eenn
an approximate $\cO(1/r)$ scaling should appear in near-critical modes in the same limit, $L \to \infty$. This is 
observed for a set of numerical simulations in which $L=16\pi$. Also plotted on Figure \ref{fig:fourier_scaling} is the
log-variance of critical (at $k=0$ for GL and $k=8$ for SH) and near-critical Fourier modes in the larger domain, and
the $\cO(1/r)$ scaling is observed for the closest-to-critical modes.\medskip

\begin{figure}[htbp]
	\psfrag{a}{(a)} 
	\psfrag{b}{(b)}
	\psfrag{r}[c]{\scriptsize $\sigma$}
	\psfrag{u}[c]{\scriptsize $r_{trans}$}
	
	\psfrag{G}[c]{\scriptsize GL}
	\psfrag{S}[c]{\scriptsize SH}
	
	\centering
	\epsfig{file=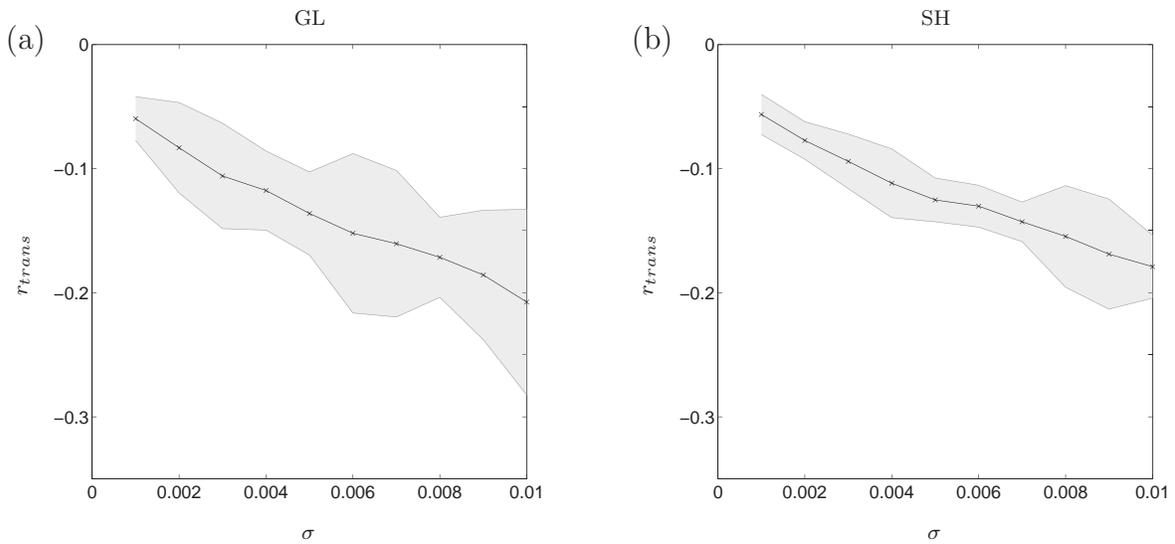,width=1\textwidth}
	\caption{Dependence of $r_{trans}$ values (where the critical Fourier mode variance diverges from an $\cO(1/r)$
scaling law) on $\sigma \in [0.001, 0.01]$ in the (a) GL and (b) SH SPDEs; the mean values of $r_{trans}$ over 10
simulations are plotted in black, and the grey regions show neighborhoods of three sample standard deviations about
these means. We take $L=2\pi$, $\xi(x,t)$ to be space-time white noise and $F(u)=1$.}
	\label{fig:linear_regime}
\end{figure}

We remark that in Figure \ref{fig:fourier_scaling}, an $\cO(1/r)$ scaling only applies 
to a parameter regime sufficiently far from $r=0$. We interpret this as the regime in which 
linearization is valid, and infer that nonlinearity is important near $r = 0$. For both the 
GL and SH equations, it is expected that these two regimes (of linear and nonlinear behavior) 
are separated by a narrow transition regime of weakly nonlinear behavior, and it can be 
shown that this weakly nonlinear regime occurs near $r = r_{trans} \sim \cO(|A|^2)$, where 
$A$ is the amplitude of the critical Fourier mode~\cite{CrossHohenberg, CrossGreenside}. Since 
$|A|$ in this stochastic setting is directly influenced by the the magnitude of the input noise 
$\sigma$, we expect that $r_{trans}$ occurs at a smaller value for smaller values of $\sigma$, {i.e.}, 
as $\sigma \to 0$, $r_{trans} \to 0^{-}$. We confirm this numerically using simulations of the GL and 
SH equations, taking $\sigma \in [0.001, 0.01]$, $r \in [-1, 0]$, and $L = 2\pi$. In each set of 
simulations using a constant value of $\sigma$ and over a range of $r \in [-1,0]$, a point $r_{trans}$ 
is computed where the critical Fourier mode variance $V_k$ first appears to diverge from the $\cO(1/r)$ 
scaling law. These $r_{trans}$ points are plotted as functions of $\sigma$ in Figure \ref{fig:linear_regime}, 
and it appears, as predicted, that $r_{trans} \to 0^{-}$ as $\sigma \to 0$. Hence, the influence of 
nonlinearity in the GL and SH SPDEs on the warning signs presented in earlier sections diminishes 
as $\sigma \to 0$.\medskip

\subsection{Other early-warning signs}
\label{ssec:ews}
Here, we describe other generic early-warning signs that have been proposed for spatio-temporal 
processes and explore them numerically in the SH SPDE. One natural measure to consider is the 
variance of $u$, which we compute by averaging spatial variance over time:
\be
	V(r) = \frac{1}{N M} \sum_{n,j=1}^{N,M} \left(u(x_j,t_n;r) - 
	\frac{1}{N} \sum_{j=1}^N u(x_j,t_n;r)\right)^2.\label{eq:spatial_var}
\ee
We note that $V(r)$ is approximately equal to the temporal variance averaged over space,
\benn
	V(r) \approx \frac{1}{N M} \sum_{n,j=1}^{N,M} \left(u(x_j,t_n;r) - 
	\frac{1}{M} \sum_{n=1}^M u(x_j,t_n;r)\right)^2,
\eenn
since $\frac{1}{N} \sum_{j=1}^N u(x_j,t;r) \approx \frac{1}{M} \sum_{n=1}^M u(x,t_n;r) \approx 0$ 
as long as $M,N$ are sufficiently large. This example suggests how temporal and spatial variance 
can be related as early-warning signs in general. In analogy to the univariate early-warning sign, 
we also compute the autocorrelation as a function of time lag $l$, averaged over space:
\be
	R_l(r) = \frac{1}{N} \sum_{j=1}^N \frac{\sum_{n=1}^{M-l} \left(u(x_j,t_n;r) - 
	\bar{u}(x_j;r)\right) \left(u(x_j,t_{n+l};r) - 
	\bar{u}(x_j;r)\right)}{\sum_{n=1}^{M} (u(x_j,t_n;r) - \bar{u}(x_j;r))^2},\label{eq:autocorr}
\ee
where $\bar{u}(x;r) = \frac{1}{M} \sum_{n=1}^M u(x,t_n;r)$. Another measure not typically considered as an
early-warning sign is the supremum of the process over all space and time:
\be
	\cS(r) = \sup_{t\in [0,T]} \sup_{x \in \cI} |u(x,t;r)|.\label{eq:supremum}
\ee
This measure generally depends on the end time, $T$, since large deviations occur as rare events. We may expect (as
with white noise) that $\cS(r) \to \infty$ as $T \to \infty$, but the rate at which certain maxima or minima of the
stochastic process diverge as $T \to \infty$ may be different for different values of $r$.\medskip

To explore the relationship between these measures and the factors of domain size, noise type, and noise correlation
length, we numerically simulated the SH SPDE for values of $r \in [-1,0]$. For some of these simulations, alternative
domain sizes of $L=2\pi$ and $L=16\pi$ are considered. For other simulations, $\xi(x,t)$ is generated as either
space-time white noise or noise that is colored in space, {i.e.}~$\E[\xi(x,t)\xi(y,s)] = C_{\textnormal{spa}}(x,y)
\delta(t-s)$. The form $C_{\textnormal{spa}}(x,y) = \exp(-(x-y)^2/\eta)$ was chosen, with $\eta = 1/32$ representing a
short correlation length and $\eta = 1/8$ representing an intermediate correlation length for domain size $L=2\pi$.
Details about the generation of space-colored noise are described in \ref{sec:gen_col_noise}. Numerical parameter
values and simulation details are otherwise as previously described.\medskip

\begin{figure}[htbp]
	\psfrag{r}[c]{\scriptsize $r$}
	\psfrag{s}[c]{\scriptsize $\log_{10}(-r)$}
	\psfrag{t}[c]{\scriptsize $r$}
	
	\psfrag{u}[c]{\scriptsize $R_1$}
	\psfrag{v}[c]{\scriptsize $\log_{10}(V)$}
	\psfrag{w}[c]{\scriptsize $\cS$}
	
	\psfrag{a}{(a)}
	\psfrag{b}{(b)}
	\psfrag{c}{(c)}
	
	\psfrag{l}[l]{\scriptsize colored noise}
	\psfrag{m}[l]{\scriptsize white noise}
	
	\centering
	\epsfig{file=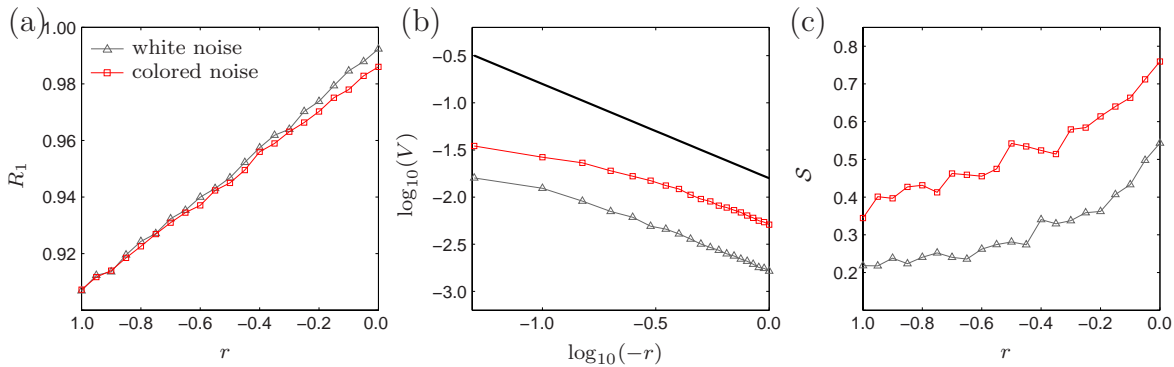,width=1\textwidth}
	\caption{Dependence of spatial and temporal statistics in solutions of the SH 
	SPDE on the parameter $r \in [-1,0]$; solutions are computed using space-time white 
	noise ($\triangle$) and spatially colored noise ($\Box$); plots show (a) autocorrelation 
	at time-lag 1 \eqref{eq:autocorr}, (b) log-log of spatial variance \eqref{eq:spatial_var}, 
	and (c) supremum of solution over all space \eqref{eq:supremum} and time. We take 
	$L=2\pi$, $F(u)=1$, $\sigma = 0.01$, and $C_{\textnormal{spa}}(x,y) = \exp(-32(x-y)^2)$ 
	for the colored noise. SPDE was numerically as described in \ref{sec:fin_diff_spde} and 
	spatially colored noise was generated approximately in Fourier space as discussed in 
	\ref{sec:gen_col_noise}.} \label{fig:diff_noises}
\end{figure}

Figure \ref{fig:diff_noises} compares the effect of white noise and spatially-colored noise on the scaling of $R_1$
\eqref{eq:autocorr}, $V$ \eqref{eq:spatial_var}, and $\cS$ \eqref{eq:supremum} with $r$. The domain size $L=2\pi$ and
the noise correlation function $C_{\textnormal{spa}}(x,y) = \exp(-32(x-y)^2)$ were used. For both types of noise, all
three measures appear to scale with $r$ in a similar way. We observe a clear monotonic increase in lag-$1$ temporal
autocorrelation as $r \to 0^-$. We also see a near $\cO(1/r)$ scaling of the spatial variance when $r$ is sufficiently
far from $r=0$. This reflects the dominant effect of critical mode variance on overall spatial variance (from Figure
\ref{fig:fourier_scaling}, we observe the variance of non-critical modes is negligible for space-time white noise).
Suprema clearly increase as $r \to 0^-$, as well. The magnitude of variation, as expressed by $V$ and $\cS$, is greater
for spatially colored noise, which we conjecture is related to the distribution of energy in the power spectrum of the
noise. The energy of space-time white noise is distributed evenly across all non-zero Fourier modes, while energy is
concentrated around the SH critical mode ($k=1$) for noise with the spatial correlation function we consider
here.\medskip

\begin{figure}[htbp]
	\psfrag{r}[c]{\scriptsize $r$}
	\psfrag{s}[c]{\scriptsize $\log_{10}(-r)$}
	\psfrag{t}[c]{\scriptsize $r$}
	
	\psfrag{u}[c]{\scriptsize $R_1$}
	\psfrag{v}[c]{\scriptsize $\log_{10}(V)$}
	\psfrag{w}[c]{\scriptsize $\cS$}
	
	\psfrag{a}{(a)}
	\psfrag{b}{(b)}
	\psfrag{c}{(c)}
	\psfrag{l}[l]{\scriptsize $C_{\textnormal{spa}} =  \exp(-8(x-y)^2)$}
	\psfrag{m}[l]{\scriptsize $C_{\textnormal{spa}} = \exp(-32(x-y)^2)$}
	
	\centering
	\epsfig{file=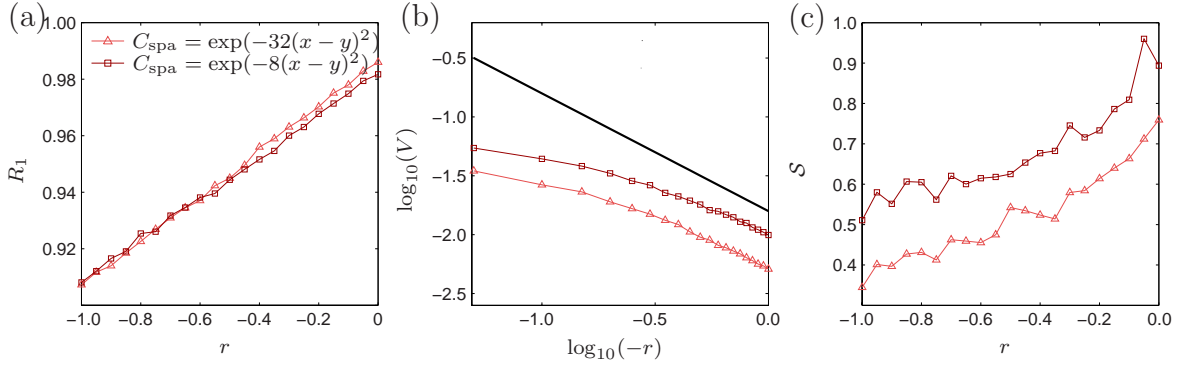,width=1\textwidth}
	\caption{Dependence of spatial and temporal statistics in solutions of the SH SPDE on the 
	parameter $r \in [-1,0]$; solutions are computed using spatially colored noise with 
	$C_{\textnormal{spa}}(x,y) =  \exp(-32(x-y)^2)$ ($\triangle$, short correlation length) and 
	$C_{\textnormal{spa}}(x,y) = \exp(-8(x-y)^2)$ ($\Box$, intermediate correlation length); plots 
	show (a) autocorrelation at time-lag 1 \eqref{eq:autocorr}, (b) log-log of spatial variance 
	\eqref{eq:spatial_var}, and (c) supremum of solution over all space and time \eqref{eq:supremum}. 
	We take $L=2\pi$, $F(u)=1$ and $\sigma = 0.01$.} \label{fig:diff_corrs}
\end{figure}

A similar comparison for two different colored noise correlation lengths is shown in Figure \ref{fig:diff_corrs}.
Again, the domain size $L=2\pi$ was used, and the noise correlation functions $C_{\textnormal{spa}}(x,y) =
\exp(-32(x-y)^2)$ (short correlation length) and $C_{\textnormal{spa}}(x,y) = \exp(-8(x-y)^2)$ (intermediate
correlation length) were considered. For both correlation lengths, all three measures once again appear to scale with
$r$ in a similar way. Additionally, we observe that the magnitude of variation is greater for the intermediate
correlation length simulations. As before, we conjecture that the reason for this can be found in the noise power
spectrum. The intermediate correlation noise has more energy in the critical SH Fourier mode than the short correlation
noise in this case.\medskip

\begin{figure}[htbp]
	\psfrag{r}[c]{\scriptsize $r$}
	\psfrag{s}[c]{\scriptsize $\log_{10}(-r)$}
	\psfrag{t}[c]{\scriptsize $r$}
	
	\psfrag{u}[c]{\scriptsize $R_1$}
	\psfrag{v}[c]{\scriptsize $\log_{10}(V)$}
	\psfrag{w}[c]{\scriptsize $\cS$}
	
	\psfrag{a}{(a)}
	\psfrag{b}{(b)}
	\psfrag{c}{(c)}
	\psfrag{l}[l]{\scriptsize $L = 16\pi$}
	\psfrag{m}[l]{\scriptsize $L = 2\pi$}
	
	\centering
	\epsfig{file=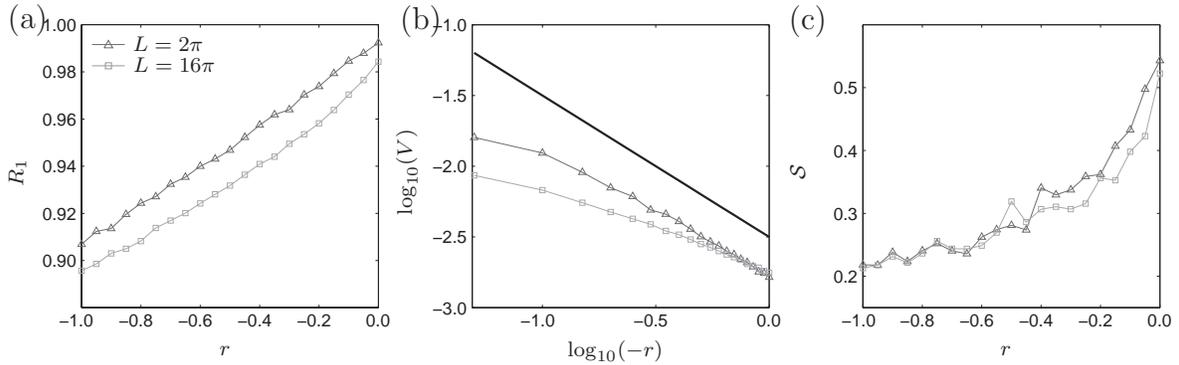,width=1\textwidth}
	\caption{Dependence of spatial and temporal statistics in solutions of the SH SPDE on the 
	parameter $r \in [-1,0]$; solutions are computed using domain sizes $L=2\pi$ ($\triangle$) 
	and $L=16\pi$ ($\Box$); plots show (a) autocorrelation at time-lag 1 \eqref{eq:autocorr}, 
	(b) log-log of spatial variance \eqref{eq:spatial_var}, and (c) supremum of solution over 
	all space and time \eqref{eq:supremum}. We take $\xi(x,t)$ to be space-time white noise in 
	both domain sizes, $F(u)=1$ and $\sigma = 0.01$.}
	\label{fig:diff_domains}
\end{figure}

Finally, Figure \ref{fig:diff_domains} shows the effect of domain size on $R_1$, $V$, and $\cS$. Space-time white noise
was used for simulations with domain sizes $L=2\pi$ and $L=16\pi$. Here, we observe that domain size has an effect on
the autocorrelation signal. Specifically, $R_1$ for $L=16\pi$ is less than $R_1$ for $L=2\pi$ for all values of $r \in
[-1,0]$. Also, we see that the larger domain size loses the overall $\cO(1/r)$ scaling in spatial variance, instead
growing more slowly as $r \to 0^-$. This could have been anticipated from the observed scaling of non-critical modes in
Figure \ref{fig:fourier_scaling}. We note that suprema appear to scale in the same way for both domain sizes, which
suggests a potential domain size invariant early-warning sign.

\section{Conclusions \& outlook}
\label{sec:conclusions}

In this paper, we have begun to develop elements of a mathematical theory for early-warning signs in pattern-forming
SPDEs, specifically in the homogeneous regime before a first bifurcation point. In particular, we gave an analytical
treatment of the scaling laws for covariance operators of the linearized SPDE problem. This analysis included the
linearized GL and SH equations, which we investigated numerically in the second part of this work. In the numerical
simulations, we focused on the influence of distance to bifurcation, noise strength, noise color and domain
size.\medskip

Although we believe that our work provides a basis for the study of early-warning signs in SPDEs, many open problems
remain. We have attempted to collect several references from different fields as additional starting points for future
research. There are many natural mathematical problems that seem to be of particular interest in spatial early-warning
sign applications. For example, given a particular pattern-forming system, can we give precise estimates for different
regime sizes and early-warning sign scaling laws in these regimes for the covariance operators (and other statistical
measures) of the linearized problem in comparison to the full nonlinear problem? What role do amplitude equations for
SPDEs play in this context? How can we classify which models display warning signs for spatio-temporal patterns? From
these questions, it is clear that there are many mathematical challenges that remain to be addressed in order to
quantify the the dynamics of stochastic systems operating near instability.\medskip

\textbf{Acknowledgements:} KG thanks Mary Silber for useful conversations throughout the duration of this project, 
and gratefully acknowledges support from the NSF Math and Climate Research Network (DMS-0940262). CK thanks the 
Austrian Academy of Sciences ({\"{O}AW}) for support via an APART fellowship and acknowledges the European 
Commission (EC/REA) for support by a Marie-Curie International Re-integration Grant. CK also thanks 
Dirk Bl\"omker for inspiring discussion at the workshop ``Infinite-Dimensional Stochastic Systems: Theory and 
Applications'' (Wittenberg, January 2014). We also thank two anonymous referees for very insightful comments, which
helped to improve the manuscript.

\appendix

\section{Numerical methods}
\label{ap:numerics}

\subsection{Generation of colored noise in space}
\label{sec:gen_col_noise}

The following procedure for generating a noise field $\xi(x,t)$ that is colored in space and white 
in time is described in~\cite{GarciaOjalvoSancho}. For $\xi(x,t)$ with $x\in (-\infty,\infty)$, it can 
be shown that if
\benn
	\E[\xi(x,t)\xi(y,s)]= C_{\textnormal{spa}}(x,y)\delta(t-s).
\eenn
and the Fourier transform of $C_{\textnormal{spa}}(x,0)$ is finite, then 
\benn
	\E[\hat{\xi}(k,t)\hat{\xi}(l,s)] = \hat{C}_{spa}(k) \delta(k + l) \delta(t-s),
\eenn
where hats denote a function transformed to Fourier $k$-space:
\benn
	\hat{f}(k) = \int_{-\infty}^{\infty} f(x) e^{-\txti kx}~ \txtd x.
\eenn
This decoupling of modes in Fourier space can be exploited to construct $\hat{\xi}(k,t)$:
\be
	\hat{\xi}(k,t) = \sqrt{\hat{C}_{\textnormal{spa}}(k)} \alpha(k,t),\label{eq:fourier_noise}
\ee
where $\alpha(k,t)$ are complex random variables such that $\alpha(-k,t) = \alpha^*(k,t)$, with real and imaginary parts drawn from the normal distribution with mean 0 and $\E[\alpha(k,t) \alpha(l,s)] = \delta(k+l)\delta(t-s)$. From here, \eqref{eq:fourier_noise} can be transformed back into real space to obtain $\xi(x,t)$. The practical 
implementation of this approach in MATLAB is performed approximately for $x \in \cI$ using the Fast Fourier Transform (FFT):

\begin{verbatim}
x=linspace(0, L, N+1); x = x(1:N); %Spatial domain
Cspa=exp(-((x-L/2).^2)); %Example spatial correlation function
Ck=fft(Cspa); %Fast Fourier transform of the correlation function

%Generate anticorrelated noise field in Fourier space
randNums=randn(1,N);
alphak=sqrt(N)*[randNums(1), sqrt(1/2)*(randNums(2:N/2)+...
    1i*randNums(N/2+1:end-1)), randNums(N),...
    sqrt(1/2)*(fliplr(randNums(2:N/2)-1i*randNums(N/2+1:end-1)))];
		
xik=alphak.*sqrt(Ck);
xi=ifft(xik); %Invert Fourier transform
\end{verbatim}

\subsection{Finite difference solution of SPDE}
\label{sec:fin_diff_spde}

A finite difference method in space is used to discretize the SPDE in space to obtain SODEs as described below. Then an
implicit Euler-Maruyama scheme is used to solve \eqref{eq:SPDE} with $F(u)=1$. The SODE scheme is described
in~\cite{KloedenPlaten,Higham}. Space and time are discretized as $(x_1, x_2,..., x_{N}) = (0, \Delta x, ..., L -
\Delta x)$ and $(t_1,t_2,...,t_M) = (0, \Delta t, ..., T)$, and the discrete solution is denoted $u(x_j,t_n) = u^n_j$.
The simulations generated for this paper used numerical parameter values $\Delta x = 0.1$, $\Delta t = 2^{-4}$,
$T=4000$, and $\textnormal{tolerance} = 0.01$. Initial conditions were taken to be uniformly random in space and were
drawn from the interval $[-0.1,0.1]$. The spatial derivatives of \eqref{eq:SPDE} are discretized using central
differencing operators, {i.e.}
\beann
	\partial^2_{x} u_j &\approx& (u_{j+1} - 2 u_j + u_{j-1})/\Delta x^2\\
	\partial^4_{x} u_j &\approx& (u_{j+2} -4 u_{j+1} + 6 u_{j} - 4 u_{j-1} + u_{j-2})/\Delta x^4.
\eeann
This discretization results in a system of coupled SODEs,
\be
	\txtd u_j = f_j(\mathbf{u}) ~\txtd t + \sigma ~\txtd W_j(t), \label{eq:numerical_SODE}
\ee
where $\mathbf{u} = (u_1, u_2, ..., u_{N})$. For instance, when $f$ is defined as in 
\eqref{eq:SH_PDE}, $f_j$ is
\beann
	f_j(\mathbf{u}) &=& (r-1) u_j - 2 \left(u_{j+1} - 2 u_j + u_{j-1} \right)/\Delta x^2 \\ 
	&&- \left(u_{j+2} - 4 u_{j+1} + 6 u_{j} - 4 u_{j-1} + u_{j-2} \right)/\Delta x^4 - u_j^3.
\eeann
To satisfy periodic boundary conditions, $u_{0} = u_{N}$. The solution of 
\eqref{eq:numerical_SODE} at $t_{n+1}$ is implicitly defined by the update equation
\be
	u^{n+1}_j = u^{n}_j + \Delta t ~ f_j(\mathbf{u}^{n+1}) + \sigma~ \txtd W^n_j.\label{eq:update_eqn}
\ee
Newton's method is used to iteratively solve \eqref{eq:update_eqn}. Explicitly, 
\eqref{eq:update_eqn} is written
\benn
	G_j(\mathbf{u}^{n+1}) = u_j^{n+1} - \Delta t ~f_j(\mathbf{u}^{n+1}) - u_j^{n} - \sigma~ \txtd W^n_j = 0.
\eenn
Then the Newton iteration formula (on the index $i=1,2,...$) for $\mathbf{u}^{n+1}$ is given by
\benn 
	\mathbf{u}^{n+1}_{i+1} = \mathbf{u}^{n+1}_{i} - (\txtD G(\mathbf{u}^{n+1}_{i}))^{-1}G(\mathbf{u}^{n+1}_{i}),
\eenn
where $\txtD G(\mathbf{u}^{n+1}_i)$ is the Jacobian matrix of $G = (G_1, G_2, ..., G_{N})$ with respect to
$\mathbf{u}$, evaluated at $\mathbf{u}^{n+1}_{i}$, and $\mathbf{u}^{n+1}_1 = \mathbf{u}^{n}$. The iteration terminates
when $\| u^{n+1}_{i+1}-u^{n+1}_{i} \|_{2}$ is less than a prescribed tolerance value.

\bibliographystyle{plain}

\end{document}